# Auto-assemblies of α-Cyclodextrin and Grafted Polysaccharides:

## Crystal Structure and Specific Properties of the Platelets

**Florent Carn,[1] Sophie Nowak,[2] Ismail Chaab,[1, 3] Raul Diaz-Salmeron,[3, 4]**

**Madeleine Djabourov, [*,3] Kawthar Bouchemal[4]**

[1] Laboratoire Matière et Systèmes Complexes, UMR 7057, Université Paris Diderot,

Sorbonne Paris Cité, 10 Rue Alice Domon et Léonie Duquet, 75205 Paris, France

[2] Plateforme Rayons X – UFR de Chimie, Université Paris Diderot,

35 rue Hélène Brion, 75205 Paris Cedex 13, France

[3] ESPCI Paris, Laboratoire de Physique Thermique, PSL Research University,

10 Rue Vauquelin 75231 Paris Cedex 5, France

[4] Institut Galien Paris Sud, CNRS, Univ. Paris-Sud, Université Paris-

Saclay, Faculté de Pharmacie, 92296, Châtenay-Malabry, France

*corresponding author e-mail: Madeleine.Djabourov@espci.fr

# Abstract

Cyclodextrins (CD) are a family of oligosaccharides with a toroid shape, which exhibit a remarkable ability to include guest molecules in their internal cavity providing a hydrophobic environment for poorly soluble molecules. Recently new types of inclusions of alpha CD with alkyl grafted polysaccharide chains (pullulan, chitosan, dextran, amylopectin, chondroitin sulfate…) have been prepared which are auto-assembled into micro and nano-platelets. We report in this paper an extensive investigation of platelets with different compositions, including their reversible hydration (Thermo Gravimetric Analysis), crystalline structure (Powder x-Rays Diffraction), dimensions and shapes, (Scanning Electron Microscopy-Field Emission Gun), thermal properties, solubility and melting (Micro Differential Scanning Calorimetry). The crystalline platelets exhibit layered structures intercalating the polysaccharide backbones and CD complexes hosting the grafted alkyl chains. The monoclinic symmetry of columnar type crystals suggests a head-to-tail arrangement of the CDs. The platelets have a preferentially hexagonal shape with sharp edges, variable sizes, and thicknesses and sometimes show incomplete layers (terraces). The crystal parameters change upon dehydration. Melting temperatures of platelets in aqueous solutions exceeds 100 °C. Finally, we discuss the potential relation between the platelet structure and applications for mucoadhesive devices.

# 1.    INTRODUCTION

Cyclodextrins (CDs) are a family of oligosaccharides with a toroid shape, which exhibit a remarkable ability to include guest molecules in their internal cavity. CDs especially are able to solubilize hydrophobic molecules which penetrate in their internal cavity. CDs are considered as natural materials originating from starch and thus "non-toxic" natural products. Besides food and cosmetics, pharmaceutical and agricultural industries are the other main users of CD molecules [1]. For instance, CDs and small molecular derivatives are used as excipients in the pharmaceutical field. In recent years CDs and grafted CDs (amphiphilic CDs, CD-polymers, CD-pendant polymers, and CD-based polyrotaxanes) became building blocks for biocompatible supramolecular assemblies. These novel functional materials take advantage of the host–guest interactions, the good biocompatibility and the functionalization capacity [2]. At the macroscale, these supramolecular assemblies can take different three-dimensional shapes such as nanospheres, nanogels, micelles, and vesicles allowing additional solubilizing properties [3]. In some other cases, aggregation of the host-guest complexes leads to the formation of visible particles, the extent of aggregation increasing with CD concentration [4].

Recently, it was reported another type of autoassemblies obtained by complexation between $\alpha$-CD molecules and hydrophobically-modified polysaccharides in solution appearing as nano or micron size platelets. It was shown that such platelets could be combined with antimicrobial drug to treat efficiently mucosal infections[5,6].

We consider in this paper the structure and specific properties of the complexes between hydrophobically-modified polysaccharides and $\alpha$-CD molecules. These nano or micrometric assemblies oberved by Transmission Electron Microscopy (TEM) appear as flat particles with geometrical shapes suggesting that the particles have a crystalline organisation. Hydrophobic groups from organic acids with variable alkyl chain lenghts were grafted on the polysaccharide chains, making the grafted polymers insoluble in water; Mixing solutions with 1% polymer and 5 to 10% $\alpha$-CD by weight in water (minimum required for complete complexation) under mild agitation, leads to the formation of a white homogeneous suspension of micro or nano size flat particles. Some of these systems have therapeutic applications. Compared with classical drug delivery systems, it appears that one of the specific chractersitcs of these self-assemblies is, at first view, their shape which contrasts with the most common spherical shapes [7,8]. The shape of microplatelets composed of polymer, alkyl chains and mostly $\alpha$-CD molecules, is necessarily related to the process of crystal nucleation and growth in solution. Consequently, the structure of these CD based aggregates is different from those reported in the literature so far. It is important therefore to fully characterize these autoassemblies and highlight their specific features and possible applications.

The pioneer investigations of Saenger and coworkers identified a variety of crystalline strucures with CD molecules and their inclusions depending on the guest molecules. The various methods of synthesis of inclusion complexes were summarized in [9]. In some cases, single cystals were obtained by slowly cooling supersatured solutions [10] facilitating a full characterization of the crystal structure down to atomic levels.

It is the aim of this investigation to explore structure, stability and solubility properties for the nano or micrometric assemblies (platelets) formed with hydrophobically modified polysaccharides and $\alpha$-CD. The role of the different components should be clarified; the stability of the auto assemblies formed without covalent bonds between constituents must be examined in aqueous environments. These investigations aim to predict the interplay between the platelets properties and their applications. We conducted a preliminary investigation on $\alpha$-CD crystals formed in supersaturated aqueous solutions at room temperature, in order to establish afterwords comparisons between the platelets and pure $\alpha$-CD crystals [11]. In the present investigation were varied both the polysaccharides (chitosan, dextran, pullulan, amylopectin, chondroitin sulfate) and the type of the grafted alkyl chains. Preparation of all the complexes followed identical protocols. The experimental techniques include powder XRD (X rays diffraction),  TGA (Thermo Gravimertic Analysis) for analysis of the hydration and SEM-FEG (Scanning Electron Microscopy with Field Emission Gun) images for the structure and shape of the platelets, and microDSC (micro Differential Scanning Calorimetry) for their thermal stability and solubility.

## 2.    EXPERIMENTAL SECTION

**Materials.** Syntheses of the different grafted polysaccharides were carried out in Institut Galien, University Paris-Sud, according to previously reported procedures. The dextran, pullulan, amylopectin grafted with palmitic acid ($CH_3(CH_2)_{14}COOH$) were synthesized as described in [12]. The chondroitin sulfate grafted with dodecyl ($CH_3(CH_2)_{11}NH_2$), hexadecyl ($CH_3(CH_2)_{15}NH_2$) or octadecylamine ($CH_3(CH_2)_{17}NH_2$) were synthesized as described in [5]. The chitosan grafted with palmitic acid, stearic ($CH_3(CH_2)_{16}COOH$) or oleic acids ($CH_3(CH_2)_7CH=CH(CH_2)_7COOH$) were synthesized as reported in [6]. A summary of the different polysaccharide backbone/hydrophobic graft combinaisons and their structures are respectively given in the Table 1S and Figure 1S of the Supporting Information (S.I.). The degree of substitution, between 2.5 to 7%, depends on the alkyl chains and on the polymer.

The α-CD sample, produced by Wacker Chemie AG, (purity ≥ 98%), was purchased from Sigma Aldrich and used as received. Milli®Q water (R = 18 MΩ cm) was used throughout the work.

**Platelets Preparation.** α-CD powder was dispersed in pure water (~ 10wt%) together with 1wt% of grafted polysaccharides under magnetic agitation at 500 rpm, at room temperature during 3 days [12]. We obtained a white homogenous suspension of platelets. By performing centrifugation with Eppendorf 5702RH centrifuge at 3000 rpm during 30 min at 25 °C a dense paste of platelets was collected. Prior to TGA analysis, the pastes were equilibrated in three desiccators at relative humidity (RH) of 94%, 32% or 3% on glass slides during 6 days, at room temperature, until they reached a constant weight. The RH in desiccators was controlled by supersaturated salt solutions (RH = 94% and 32%) and dry silica gel (RH = 5%).

**Thermogravimetric Analysis (TGA)** was performed with Q5000 set-up, from TA Instruments, on platelets with various compositions (by adapting the protocol of [11]). Samples were placed in an open platinum pan (100 µL) hung in the furnace. The initial weight of platelets was around 30 mg and nitrogen was used as the purge gas at a fixed flow of 25 mL/min. The weight of material was recorded during heating from 20 to 150 °C at a heating rate of + 1 °C/min. The heating ramp started immediately after the sample was placed on the platinum pan.

**X-Rays Diffraction (XRD)** patterns were recorded with a Panalytical Empyrean diffractometer, equipped with a multichannel detector (PIXcel 3D) and a Cu Kα X-ray source (λ = 1.54 Å) in a Bragg-Brentano θ-θ configuration. The measurements were performed in the 2.5 °– 30 ° range, with a scan step of 0.013 ° for 60 s. Simulated spectra were obtained using the Highscore Plus 3.0® software, allowing determination of the space group by Highscore plus 3.0 and FullProf Suite Program 3.0. In the software the most probable unit cell for each spectrum is searched by dichotomy, by varying the values of the lattice parameters (i.e. Dicvol method). XRD measurements were done on the paste of platelets saturated by water collected after the centrifugation of suspensions and after paste drying in an oven at 60 °C during 40 min.

**Scanning Electron Microscopy With Field Emission Gun (SEM FEG).** Images were obtained with a Zeiss Merlin Compact microscope operating at 10 kV. Before observations, the stock dispersions were stirred with vortex and diluted by a factor ten with Milli®Q water. 10 µL of these diluted dispersions were let to dry directly on a conductive double face adhesive carbon tape. The samples were carbon coated before observations.

**Microcalorimetry** (µDSC) experiments were performed with µDSC3evo from Setaram (Caluire, France) in batch Hastaloy cells by weighing as prepared suspensions of different platelets. The total weight of the samples was around 0.8 g. The reference cell was filled with water. The temperature ramp started at 25 °C after thermal equilibration and the heating was performed with a constant rate of + 0.1 °C/min to a final temperature of 120 °C.

# 3. RESULTS AND DISCUSSION

### 3.1 Composition of the Platelets.
The platelets are composed of $\alpha$-CD molecules, polysaccharides, grafted alkyl chains and water. Alkyl chains thread inside the $\alpha$-CD cavity (height of the torus *circa* 7 Å) with an average ratio of 5 $CH_2$ groups per $\alpha$-CD determined for organic acids [13]. In the platelet preparation, the concentration of $\alpha$-CD molecules was adjuted so that the total amount of $\alpha$-CD molecules was incorporated in the platelets. Generaly, this was around 10%, but with chondroitin sulfate is was lower, around 5%.

TGA was employed to determine the hydration of the platelets following the experimental strategy previously established for $\alpha$-CD crystals grown in supersaturated aqueous solutions [11]. The crystals grown in supersaturated aqueous solutions are hydrates with water molecules located inside the crystals with specific thermal stabilities. The water content varies after equilibration of the crystals in desiccators at different relative humidity (RH) values. In agreement with [14], it was found that the $\alpha$-CD crystals contain, at the most, 6 $H_2O$ molecules per $\alpha$-CD and that 2 $H_2O$ molecules are located inside the CD cavity. In the same manner, we show here that TGA allows an accurate determination of the amount of water molecules present in the platelets with various polymer and alkyl chain compositions and RH values. Figure 1 represents some typical examples of TGA scans after sample equilibration. Figures 1 "A" show the percentage of weight loss from initial weight versus temperature for: palmitoyl amylopectin/$\alpha$-CD, palmitoyl dextran/$\alpha$-CD and palmitoyl chitosan/$\alpha$-CD equilibrated at different RH values. The curves in Figures "B", are the derivatives of the curves from Figures 1"A".

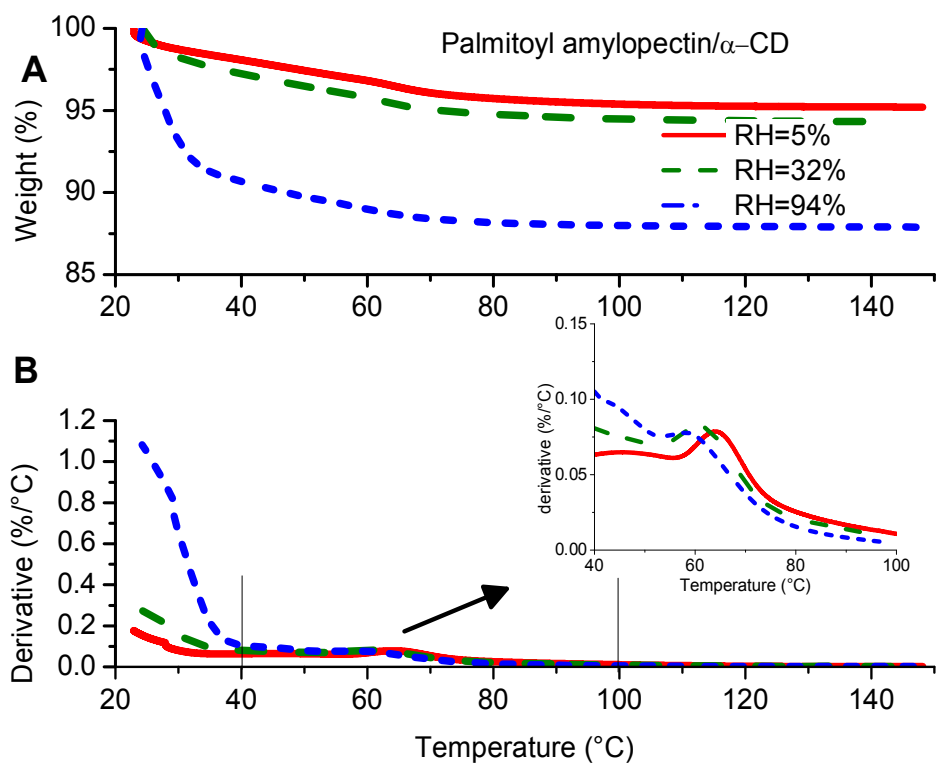

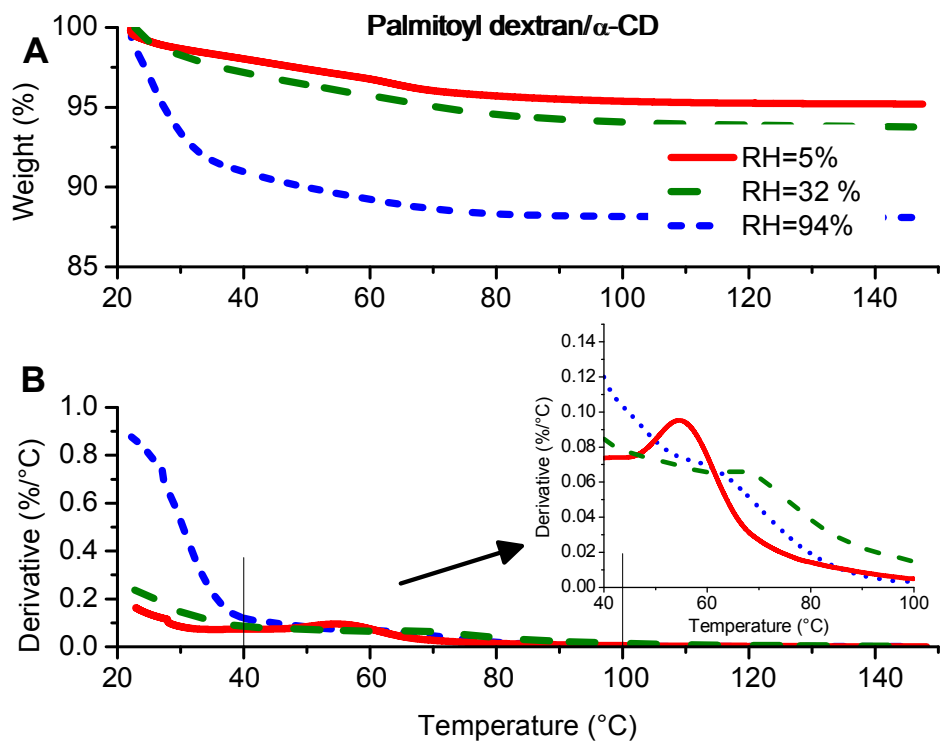

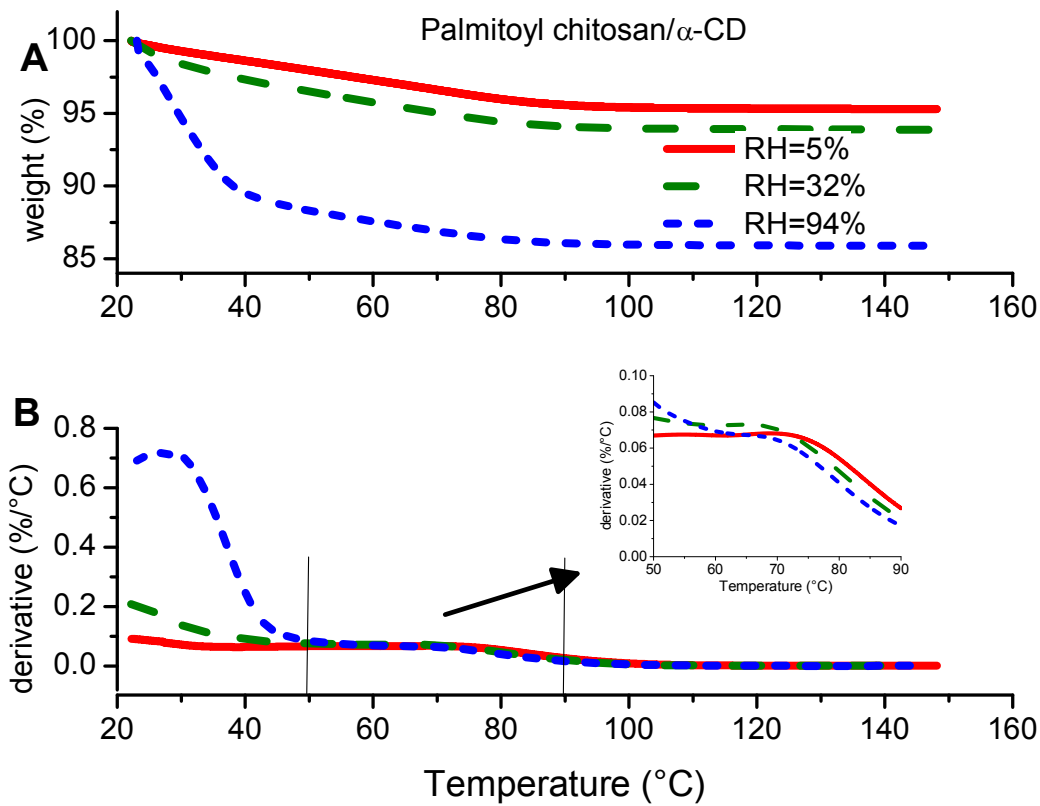

**Figure 1**. TGA scans -from top to bottom- of: palmitoyl amylopectin, dextran and chitosan/α-CD platelets previously equilibrated under different RH indicated on the figure. The heating ramp is +1 °C/min: (**A**) Mass loss in wt.% versus temperature. (**B**) Derivative of the mass loss versus temperature; the vertical lines delimitate the dehydration peak of the "structural water". The inset in figures B represents the mass loss derivative with temperature in the domain of dehydration of the "structural water".

The platelets equilibrated at different RH retain variable amounts of water, increasing with RH. During the heating ramps, between 20 and 120 °C, water molecules desorb continuously from the platelets. From the derivative curves, one can define two states of water in the platelet structure: i) water that desorbs between 20 and 40 °C, with a continuously decreasing rate; ii) a small fraction of water that desorbs between 40-50 and 100 °C with a broad maximum, showing a high thermal stability. We call this fraction "structural water". Table 1 summarizes the hydration properties of different platelets including oleoyl chitosan/$\alpha$-CD and stearoyl chitosan/$\alpha$-CD evaluated by weight of dry material.

**Table 1. Water Content of Platelets Composed of Hydrophobically Modified Polysaccharides and $\alpha$-CD versus Relative Humidity (RH): Total Water and Structural Water wt.% Dry Solid**

| Platelet composition | Hydration (wt.% water/dry | RH = | RH = | RH |
|---|---|---|---|---|
| Oleoyl chitosan/H₂O/$\alpha$-CD | Total water | 13.0 | 7.6± | 5.4 |
| | *Structural water (±0.06)* | *3.3* | *2.5* | *2.2* |
| Stearoyl chitosan/H₂O/$\alpha$-CD | Total water | 13.7 | 6.8± | 4.7 |
| | *Structural water (±0.06)* | *3.2* | *2.9* | *2.7* |
| Palmitoyl chitosan/H₂O/$\alpha$-CD D | Total water | 14.1 | 6.1± | 4.6 |
| | *Structural water (±0.06)* | *3.3* | *3.1* | *3* |
| Palmitoyl dextran/H₂O/$\alpha$-CD | Total water | 13.6 | 6.3 ± | 4.8 |
| | *Structural water (±0.04)* | *2.9* | *3.0* | *2.8* |
| Palmitoyl pullulan/H₂O/$\alpha$-CD | Total water | 13 ± | 6.3 ± | 4.8 |
| | *Structural water (±0.02)* | 3.4 | 3.2 | 2.8 |
| Palmitoyl amylo/H₂O/$\alpha$-CD | Total water | 11.6 | 7.5 ± | 4.8 |
| | *Structural water (±0.02)* | *3.0* | *3.5* | *3.0* |

The maximum water adsorbed at RH = 94% varies between 11.6 and 14% by weight of dry material, and the minimum, at RH = 5%, between 4.8 and 5.4%. A first striking result is that "structural water" represents ≈ 3% whatever the equilibration RH and the nature of the grafted polysaccharide (nature of the polysaccharide and of the grafted chain).

In the complexes, the $\alpha$-CD cavity is filled with the grafted alkyl chains and water is unable to occupy these sites, contrary to the case of pure $\alpha$-CD hydrated crystals. Adsorbed water molecules are now located in the interstitial positions between $\alpha$-CD molecules and around the backbone of the polysaccharides. The dry material of platelets is mainly composed of $\alpha$-CD molecules, with a small proportion, by weight about ~ 10%, of polymers, according to the preparation conditions. Water molecules are shared between: a) the crystalline part, containing $\alpha$-CD molecules and included grafted chains, and b) the polysaccharide backbones, which are hydrophilic, but the precise reparation is unknown: water sorption isotherms of polysaccharide films (the backbones) are reported in the literature and vary with the macromolecule. However, there are additional difficulties to evaluate water repartition in platelets: in our systems, we could not evaluate the exact weight of backbone, compared to total weight of polymer; the confinement of the polymer inside the platelet structure might also modify water adsorption at various RH values. With a crude approximation, ignoring the water molecules adsorbed on the polysaccharide backbones, we estimate that the number of the water molecules in the most hydrated crystals should not exceed 7-8 molecules/$\alpha$-CD.

The amount of "structural water" is constant for all the samples (and is much smaller than the total water content), suggesting that the structural water is interstitial water in the CD crystals. This is further supported by the fact that the structural water is harder to remove, which would be expected if it is in the CD crystals, due to the extensive hydrogen bonding that exists between water and CD molecules.

### 3.2 Crystal Structure

*3.2.1 X-ray Diffraction Patterns.* We investigated all samples of platelets in the humid state, by collecting the sediments from the bottom of the falcon tubes after centrifugation. The diffraction patterns were recorded within 600 s at 25 °C between $2\theta=2$ and 30 °. We checked by performing two successive measurements the absence of structural evolution on this time scale due to drying under the X ray beam. Figure 2 shows the comparison of the diffraction patterns of chitosan grafted with different alkyl chains: stearoyl, palmitoyl and oleoyl. One can notice first that the patterns are identical in this angular domain and exhibit numerous narrow peaks. Some samples exhibit a higher degree of crystallinity (stearoyl, palmitoyl). These patterns are different from pure $\alpha$-CD hydrated crystals that are in the "cage" type structure [11]: in particular, there is a distinct diffraction peak at $2\theta\approx20$ °. Hwang *et al* [15] working with PEG/$\alpha$-CD inclusion complexes noticed that the strong reflection at $2\theta\approx20$ ° is characteristic of $\alpha$-CD molecules involved in a "channel" type structure and is missing in the "cage" structure. The associated interreticular distance, $d_{2\Theta=20°}$ = 4.4 Å corresponds to the diameter of the internal cavities of $\alpha$-CDs which constitute the channels when the $\alpha$-CD molecules are threaded along a polymer chain. The whole curve reported by [15] is however different from ours. We can consider that the inclusion complexes formed with grafted chitosan are quite similar, in the three cases shown in Figure 2.

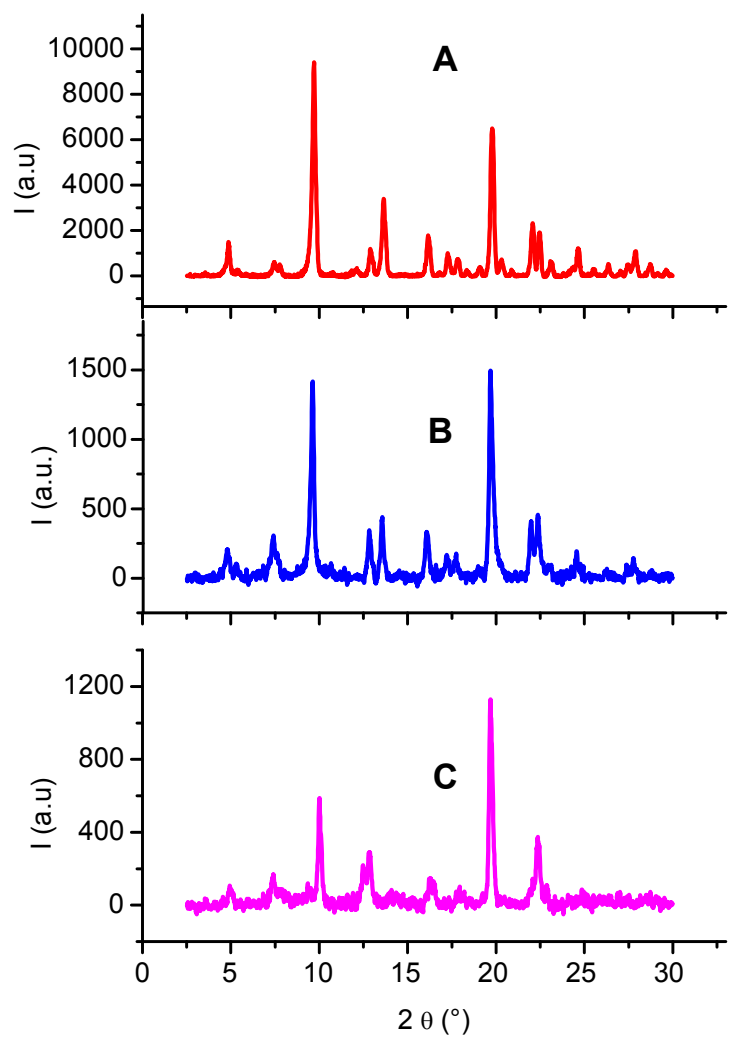

**Figure 2**. X-Ray diffraction patterns of humid platelets of inclusion complexes based on chitosan: A) stearoyl chitosan/α-CD, B) palmitoyl chitosan/α-CD and C) oleoyl chitosan/α-CD.

Figure 3 shows the X-Ray diffraction patterns of inclusion complexes in platelets with different polysaccharides (dextran, pullulan, amylopectin) grafted with palmitoyl and of chondroitin sulfate grafted with hexa and octadecyl amine. The crystallinity of the platelets is mostly visible with the palmitoyl grafted polymers, including chitosan (Figure 2) and less obvious with the grafted dodecylamine. The hydrophobic groups and degree of substitution control the growth of the crystals. Two large peaks appear in all the spectra at 2θ≈9.6 ° and 2θ≈19.7 °. The experimental results demonstrate that the patterns are almost identical in all the types of humid platelets that we investigated.

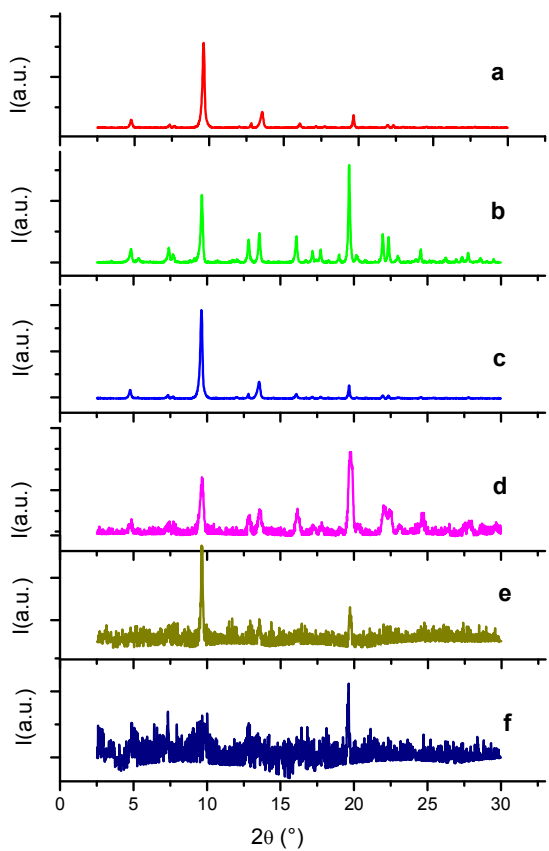

**Figure 3**. X-Ray diffraction patterns of humid platelets of inclusion complexes of palmitoyl - a) dextran/α-CD; b) amylopectin/α-CD; c) pullulan; d) hexadecylamine chondroitin sulfate /α-CD; e) octadecylamine chondroitin sulfate /α-CD; f) dodecylamine chondroitin sulfate /α-CD.

Figure 4 shows the diffraction pattern of palmitoyl amylopectin/α-CD inclusion complex after drying the humid paste during 40 min at 60 °C. The platelets with a high degree of crystallinity keep their organized structure upon drying, but the peaks shift, indicating that the structure rearranges, after the sites occupied by water molecules become vacant. These changes illustrate the "adaptability" of the crystals to adjust their volume to the hydration/dehydration process. It is important to highlight that the platelets contain also polymer backbones, which are included in the crystals in interstitial positions; they do not penetrate inside the cavities. The presence of water molecules in the interstices of the crystals allows the polymer to remain in a hydrophilic environment in humid platelets.

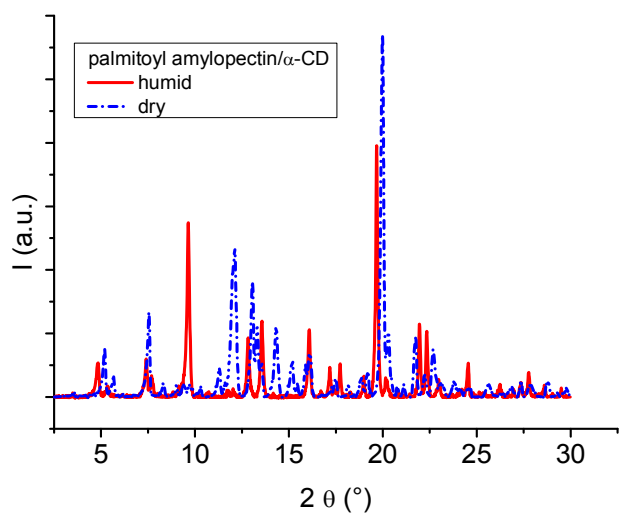

**Figure 4**. Comparison between X-ray patterns of high crystallinity platelets of humid and dry palmitoyl amylopectin /α-CD.

*3.2.2 Interpretation of Diffraction Patterns.* We report on Table 2 the position of the maxima of diffraction peaks observed in Figures 2 and 3 for hydrated platelets. It appears that these positions are almost identical with a standard deviation of 0.03 ° whatever the nature of the polymer. We decided to analyze the diffraction pattern of palmitoyl amylopectin/α-CD humid sample (Figure 3b) as the best-resolved example of the structure of platelets by using Highscore Plus 3.0® software, allowing determination of space group and of unit cell parameters. Figure 5 shows the comparison between the simulated and measured patterns using the "size-strain analysis" option, which also takes into account the broadening of the peaks. One can see a very good agreement over the entire angular range. The figure indicates also the Miller indexes. Table 2 reports the peak positions of all samples and at the bottom line, of the theoretical pattern.

**Table 2.** $2\theta_{max}$ (°) of the Main Diffraction Peaks in the Patterns Shown in Figures 2 and 3
We report the positions of nine successive peaks. In the bottom line are the theoretical positions for these peaks.

| Polym/Graft | $2\theta_{max}$ (°) Peak positions | | | | | | | | |
|---|---|---|---|---|---|---|---|---|---|
| Dextr/Palm | 4.83 | 7.39 | 9.64 | 12.82 | 13.56 | 16.08 | 19.67 | 21.95 | 22.34 |
| Pullul/Palm | 4.85 | 7.39 | 9.66 | 12.82 | 13.54 | 16.08 | 19.71 | 21.95 | 22.37 |
| Amylo/Palm | 4.83 | 7.40 | 9.64 | 12.84 | 13.57 | 16.08 | 19.67 | 21.96 | 22.33 |
| Chon/Hexa | 4.77 | 7.36 | 9.62 | . | 13.54 | 16.07 | 19.67 | . | . |
| Chon/Octa | . | . | 9.62 | 12.80 | 13.54 | . | 19.71 | . | . |
| Chon/Dod | . | 7.39 | 9.59 | . | . | . | 19.71 | . | . |
| Chito/Palm | 4.79 | 7.41 | 9.62 | 12.83 | 13.56 | 16.08 | 19.68 | 20.77 | 22.36 |
| Chito/Stea | 4.86 | 7.43 | 9.71 | 12.88 | 13.65 | 16.14 | 19.78 | 20.86 | 22.45 |
| Chito/Oleo | - | - | 9.61 | 12.83 | 13.63 | 16.21 | 19.69 | 20.89 | 22.36 |
| Model | 4.81 | 7.40 | 9.63 | 12.83 | 13.56 | 16.08 | 19.66 | 21.95 | 22.33 |

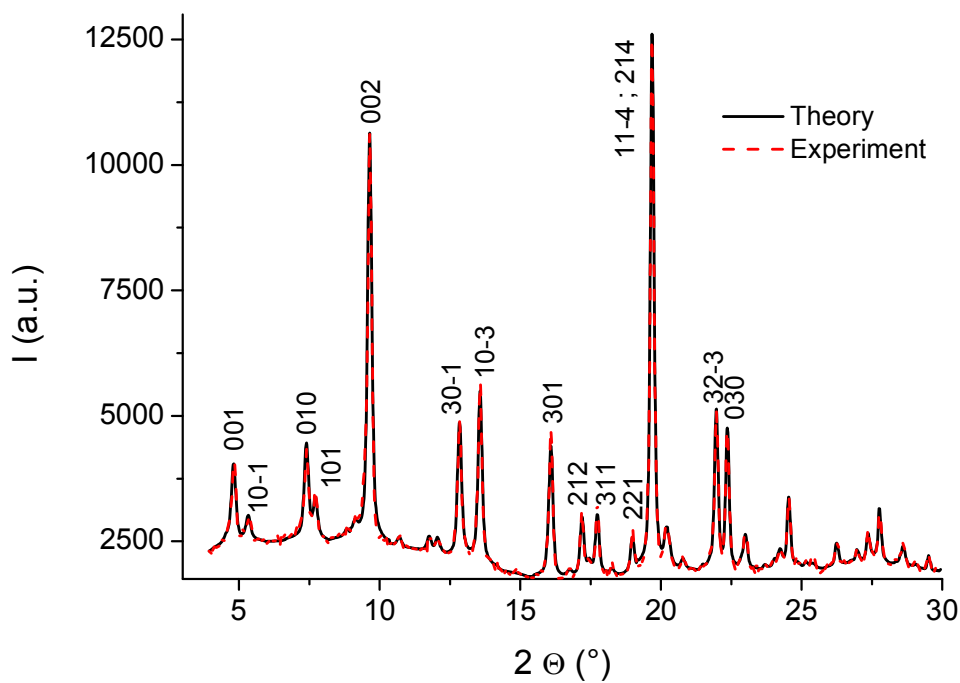

**Figure 5**. Comparison between X-ray diffraction patterns simulated (in black solid line) and measured with humid platelets of pamitoyl amylopectin/α-CD (in red dash line). Miller indexes are indicated on the peak summit.

The simulation determines the unit cell parameters and the space group symmetry. In each case all the peaks match, and the lattice parameters and the space group are chosen with the best figure of merit. We analyzed in same way the patterns of several dried samples. Table 3 reports the unit cell parameters for humid platelets and different dried platelets. Table 4 gives explicitly the diffraction peaks in humid and dry palmitoyl amylopectin/α-CD samples.

**Table 3. Crystallographic Parameters Characterizing the Structure of Platelets: Symmetry, Space Group, Unit Cell Lengths (*a*, *b*, *c*) and Angles (α, β, γ) and Unit Cell Volume**

| sample | symmetry | Space group | *a* (Å) | *b* (Å) | *c* (Å) | α (°) | β (°) | γ (°) | Vol (Å$^3$) |
|---|---|---|---|---|---|---|---|---|---|
| **Humid platelet**s | monoclinic | P121 | 20.7 | 11.9 | 19.5 | 90 | 110.4 | 90 | 4526 |
| **Dried Pullul Palm/α-CD** | triclinic | P1 | 16.6 | 12.0 | 9.3 | 87.9 | 99.0 | 89.9 | 1847 |
| **Dried Amylo Palm/α-CD** | monoclinic | P121 | 19.1 | 11.7 | 18.6 | 90 | 112.6 | 90 | 3872 |
| **Dried Dext Palm /α-CD** | monoclinic | P121 | 19.1 | 11.7 | 18.7 | 90 | 112.7 | 90 | 3873 |

**Table 4. Comparison between 2θ$_{max}$, Distances and Miller Indexes of Humid and Dried Palmitoyl Amylopectin /α-CD Platelets**

| Humid platelets | | | Dried platelets | | |
|---|---|---|---|---|---|
| **2θ$_{max}$** | ***d* (Å)** | ***hkl*** | **2θ$_{max}$** | ***d* (Å)** | ***hkl*** |
| 4.83 | 18.2 | 001 | 5.14 | 17.2 | 001 |
| 7.40 | 11.9 | 101 | 7.51 | 11.7 | 010 |
| 9.64 | 9.1 | 002 | 10.28 | 8.6 | 002 |
| 12.84 | 6.9 | 30-1 | 11.98 | 7.3 | 21-1 |
| 13.57 | 6.5 | 10-3 | 12.16 | 7.2 | 11-2 |
| 16.08 | 5.5 | 30-3 | 13.06 | 6.7 | 102 |
| 19.67 | 4.5 | 203; 21-4 | 13.54 | 6.5 | 21-2 |
| 21.96 | 4.0 | 32-3 | 14.28 | 6.2 | 10-3 |
| 22.33 | 3.9 | 123 | 19.87 | 4.4 | 221 |
| | | | 19.98 | 4.4 | 122 |

The unit cell parameters *a* and *c*, of the humid samples, are larger than the diameter of the α-CD ring, around 15 Å and *b* is larger than the height of the α-CD ring, around 7 Å. Although our analysis on a polycrystalline paste does not enable to define precisely the position of the α-CD rings inside the unit cells, we can make some remarks on the relative volume of the unit cell compared to the one of hydrated α-CD. The volume of one α-CD molecule approximated by a cylinder is close to 1240 Å$^3$ (diameter 15 Å and height 7 Å). We established that several water molecules are present in the humid platelets sitting outside the cavities and filling the space between CD molecules. Therefore, the unitary "brick", which constitutes the platelets, is around 1240 $Å^3$ in humid and totally dried platelets. In humid platelets, the volume of the unit cell is 4526 $Å^3$ and in dried palmitoyl amylopectin and dextran/$\alpha$-CD platelets, the volume of unit cell is 3872 $Å^3$. Therefore, a unit cell contains a maximum of three "bricks". The difference between the unit cell volume and the "bricks" volume correspond mainly to the volume occupied by water molecules. The dried palmitoyl pullulan /$\alpha$-CD platelets however have a unit cell of 1847 $Å^3$, which limits to one "brick" per unit cell. The lengths $a$, and especially $c$, exhibit large changes, $b$ is almost unchanged. It is important to remind that the polymer backbone is also in the interstitial sites (see below) and necessarily increases the volume between the crystalline parts like a "mortar", while the grafted alkyl chains are included in the CD cavities.

The integral breadths of the diffraction peaks, as the ratio of a peak area to its maximum, are presented in Supporting Information. The size obtained from the integral breadth is the thickness of a crystal in a direction perpendicular to the crystallographic plane producing a given reflection, neglecting contributions from other factors (instrumental factors, microstrain or chemical inhomogeneity. In humid platelets of palmitoyl amylopectin/$\alpha$-CD, the average crystal thickness is around 75 nm, it decreases to 45 nm in the dried samples. These estimations indicate that the crystalline domains are on average nanometric in their lower dimension and that a significant densification occurs upon drying.

*3.2.3 Discussion.* The platelets are supposed to belong to the category of "columnar" or "channel" type structures, due to the inclusions of polymeric chains (alkyl chains). Although the literature on X-Ray patterns of channel type structures with $\alpha$-CD molecules is well documented [16–22], the X-Ray patterns are not identical in all these studies; the control of humidity differ, many samples were thoroughly dried. Some authors followed the kinetics of the phase transformation with time resolved X-Ray diffraction. In particular, time-dependent changes in the intensity of the peak at $2\theta=20$ ° by rearrangement of the cage structure identifies the channel structure during inclusion formation [22]. In some cases, inclusion complexes were prepared as single crystals [23]. In the latter example, unit cell volume was 12 300 $Å^3$ and it was established that the $\alpha$-CD molecules are packed as dimers. Rodríquez-Llamazares *et al* [24] found that the volume of a unit cell of decanoic acid/$\alpha$-CD was 12 100 $Å^3$. Harata and Kawano [19] found for isosorbide dinitrate/$\alpha$-CD a volume of 10 389 $Å^3$. Besides, Manor and Sanger [14] determined for *n*-propanol and *n*-butanol volumes around 5 000 $Å^3$, McMullan *et al* [25] report values around 3 000 $Å^3$ in their list of different types of inclusions.

Comparing the data from different authors one can see large differences in the unit cell volumes: probably the hydration is important, but also the process of crystallization itself may lead to different arrangements. The inclusions contribute to stabilize the crystal. Harata and Kawano [19] illustrate the arrangements of a variety of channel-type structures classified according to the packing mode of $\alpha$-CD molecules, either head-to-head or of head-to-tail and to hydrogen bonds formation. They show in particular, that in the head-to-tail structure, $\alpha$-CD rings are perpendicular to the column axis in orthorhombic crystals, while their molecular axis is inclined against the column axis in monoclinic crystal type. Channel-type structures are guest-modulated, so are dimensions and angles of the unit cells. According to these authors, the monoclinic symmetry corresponds to the head-to-tail type arrangement, where CD molecules tilt against the column axis. This might be the packing mode in humid, and in two types of our dried platelets.

### 3.3 Crystal Size and Morphology. The shape and size of the platelets is one of the major concerns in applications. Platelets larger than 1 μm are non-Brownian particles; in suspensions, they are able to sediment progressively. The aspect ratio of the micrometric

platelets is also important; large, flat and thin particles become flexible and can break under shear. SEM FEG performed on dried droplets of diluted suspensions provide rapidly images with numerous platelets without further preparation and with a good resolution. Figures 6 and 7 are images of the platelets with two different magnifications.

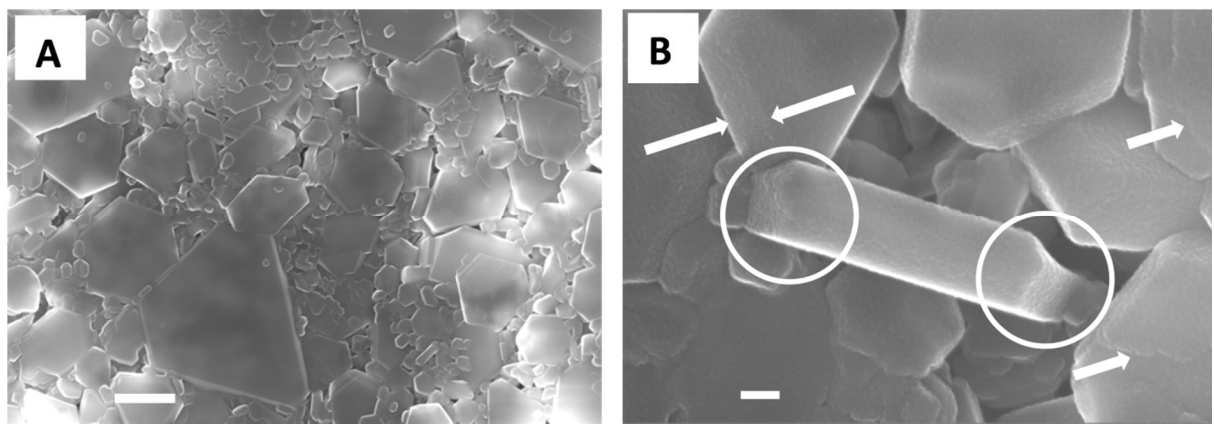

**Figure 6.** Palmitoyl chitosan/$\alpha$-CD platelets at two different magnifications: A) left side, the scale bar is 1 µm; B) right side the scale bar is 100 nm.

In Figure 6A one can see a large number of flat particles with geometric hexagonal irregular shapes and sizes. Many small hexagons ($\sim$ 300 nm) coexist with some large ones ($\sim$ 3 µm). There is a difference of, at least, one order of magnitude between the dimensions of the flat platelets in this sample. They all have sharp edges, as expected from crystalline structures. The growth of the crystals is definitely preferential in 2D. Figure 6B shows other details: the two arrows point out the thickness, around 100 nm, of a platelet with a size of 400-500 nm. Smaller platelets are thinner. On the other platelets, (individual arrows), one can see exfoliation of the flat surfaces with very thin terraces, which build layer by layer the thickness of the hexagonal particles. This interesting feature may explain the way in which the backbone of the polymer is included inside the crystals. The minimum layer thickness may be of the order of several heights of $\alpha$-CD rings, threaded along the grafted chains, (5-10 nm) with polymer backbones in between the layers. The fragment of platelet in the middle of figure 6A shows (circles) a superposition of many distinct layers at both ends.

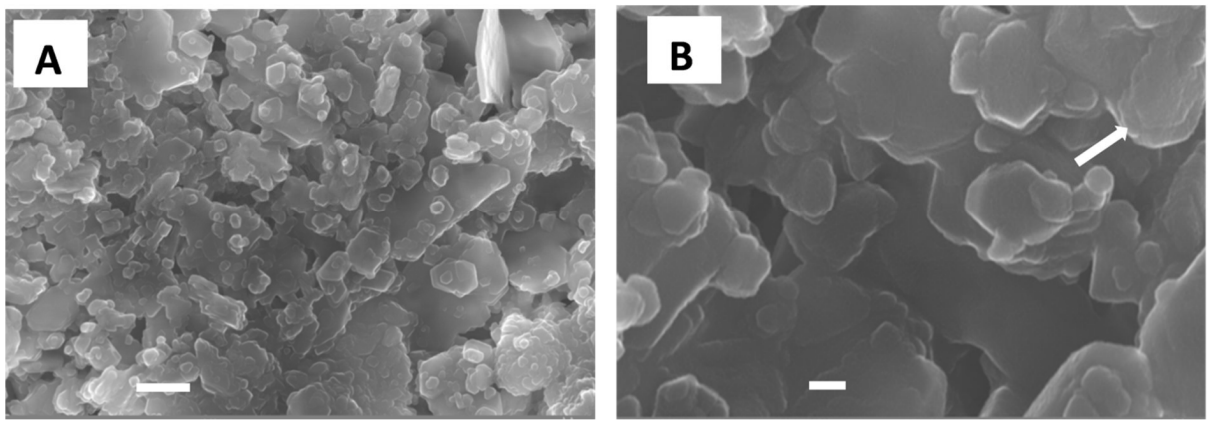

**Figure 7**. Oleoyl chitosan/α-CD platelets at two different magnifications: A) left side, the scale bar is 1 μm, B) right side, the scale bar is 100 nm.

Figure 7A and 7B show platelets of oleoyl chitosan/α-CD with the same magnifications as in Figure 6: on the left, Figure 7A, one can see platelets, which are much smaller, more homogenously distributed (100-500 nm), thinner and with more rounded edges. There is clearly an influence of the grafted chain on the growth and aspect of the platelets. On the right side, Figure 7B, one can see again, at the point of the arrow, exfoliated small layers; the hexagonal shapes are hardy visible as the rims seem damaged or worn.

The images provide a new insight on the organization of the platelets: there are major differences in the size and polydispersity of the platelets obtained with the same polymer (in this case chitosan), but different grafted chains. This suggests that the spontaneous growth rate of crystals (related to intrinsic properties of the components) has a large effect. External factors, such as the magnetic bar agitation may exert also an influence on the distribution of sizes.

To illustrate the global picture arising from this study, we propose in Figure 8, a naïve scheme summarizing both X-Ray diffraction measurements and SEM. The polymer (polysaccharide) is water-soluble and the coils anchored on the surface of the platelets (loops) can swell in contact with water. A large number of the loops exposed to water covers the surface of the platelets and creates a soft, curly, coat. The grafted alkyl chains and α-CD molecules build the hexagonal flat crystals, which include the backbones of the related polysaccharides. The platelets are sandwiched between two dense polymer layers swollen in water.

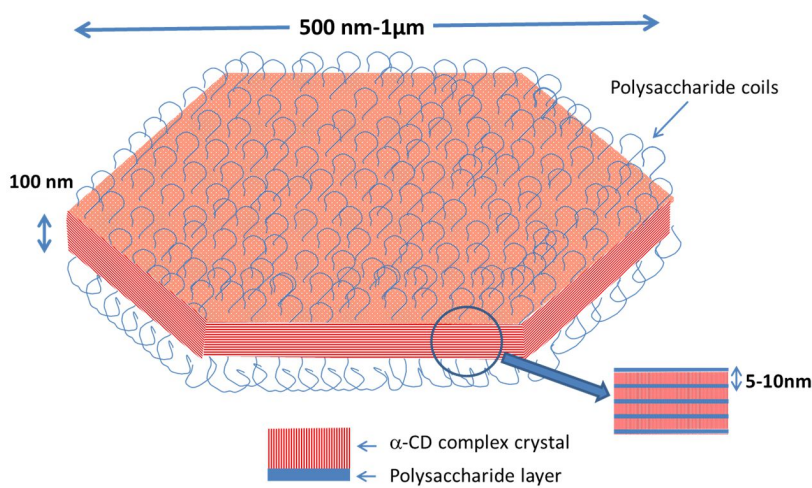

**Figure 8**. Schematic view of the platelet organization in water: it is a crystalline multi layered structure with hexagonal shape; the layers alternate CD rich and polymer rich domains; in water environment, the platelet is sandwiched between the swollen polymer coils (polysaccharides) protruding from the flat surfaces.

### 3.4 Thermal and Interfacial Properties. We examined previously [12] the thermal stability of some of the platelets suspended in Pickering emulsions and at interfaces of oil droplets.

In this paper, we investigate by μDSC the other platelets, as prepared in suspensions. We report in Figure 9 the heat flow traces of three different platelets with palmitoyl-grafted chains: chitosan, pullulan and dextran. One can see three domains: i) below 60 °C, the base lines are flat and indicate that the platelets do not dissolve; ii) between 60 and 100 °C, some minor changes appear: pullulan and dextran complexes partially dissolve (increasing slope of the heat flow) and chitosan has a small endothermal peak; iii) large sharp endothermal peaks indicate the cooperative melting transitions of the platelets above 100 °C.

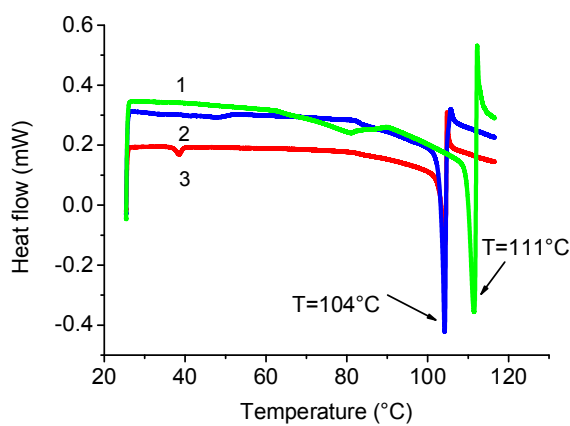

**Figure 9**. Thermal behavior of platelets in aqueous solution: the heat flow traces show the melting peaks of: (1) palmitoyl chitosan /$\alpha$-CD, at $T_m$ = 111 °C; (2) palmitoyl pullulan ; (3) palmitoyl dextran/$\alpha$-CD [12] at $T_m$ = 104 °C.

The platelets exhibit a remarkable thermal stability of the hydrated crystals in water. The melting temperatures of palmitoyl pullulan and dextran/$\alpha$-CD crystals are identical, $T_m$ = 104 °C and for palmitoyl chitosan/$\alpha$-CD, $T_m$ = 111 °C. These temperatures are above the boiling temperature of water and such measurements require a cell that supports pressures above atmospheric. Melting induces de-threading of the inclusions [12]. This remarkable stability is due to the strong hydrogen bonding of the crystals. The polymer has an influence also, on the thermal stability, as we can see from the difference of melting temperatures between chitosan and others. Other factors, which may play a role, are the molecular weights of the polymer, the ionic character or the length of the alkyl chains. Small imperfective crystals may dissolve continuously.

Investigation of interfacial and thermal properties of Pickering emulsions containing dextran, pullulan and amylopectin shows that the smallest platelets adsorb at the droplets surface (drops of 20 μm). It is shown that the aptitude of oil molecules to be threaded in α-CD cavity is a determining parameter in emulsification and stability. This important factor allows the partial substitution by another guest molecule (oil) within the platelets, which takes place at room temperature spontaneously at the oil/water interfaces without dissolution of the platelets, possibly by a diffusion mechanism of the oil [12]. Accordingly, the dissolution and the cooperative melting temperatures of the inclusion crystals changed, showing marked differences on the type of guest molecules.

### 3.5 Can Platelets Enhance Bioadhesion?

Nanocarriers composed of functionalized CDs and chitosan are reported in the literature [7,8,26,27]. Mainly spherical nanoparticles (300-500 nm) are designed for drug delivery: drugs are first partially complexed by CDs via their hydrophobic side chains and chitosan networks are prepared by ionotropic gelation [28]. Nanocarriers release insulin or heparin by diffusion across the polymer network. Nanocarriers with smaller sizes (100-200 nm) are used for gene delivery [27]. In all these examples, chitosan constitutes the matrix forming a gel network crosslinked by ionic interactions and entrapping functionalized CDs. Other examples of spherical nanoparticles obtained by self-assembly of CD polymers and hydrophobically modified dextran are reported by Nielsen *et al* [29], they can be disrupted by molecules competing for the CD cavities.

The nano or micrometric platelets presented in this paper have a very different assembling mechanism. The grafted polymers (polysaccharide) and CDs are arranged into crystalline structures. We observed on SEM FEG images successive layers with a weak interlayer cohesion, which is one of the specific features of their structure. Our investigation shows that there is no dissolution, no melting, no swelling of the platelets dispersed in water at the body temperatures. There is no ionic crosslinking of the chitosan chains in platelets, like in the spherical nanocarriers. Nevertheless, therapeutic applications have been reported [5,6] for the nanoplatelets containing chitosan suggesting that platelet structures may enhance bioavailability of chitosan.

It is well known that upon protonation of amino groups, chitosan chains are stretched by electrostatic repulsions[30]. The pH influences the conformation of chitosan chains. We suppose that the chitosan molecules anchored at the surface of the platelets are positively charged and stretched. The strong interaction between chitosan and mucin at different pH values (pH = 1 and 5.5) in solutions has been proved experimentally *in vitro* and explains the bioadhesive properties of chitosan[31]. The adhesion force increases with chitosan concentration. Bioadhesive drug delivery devices aim to be localized on biological surfaces and the bioadhesive force between the device and the biological surface is required to retain the device. Many characteristics of the polymers contribute to this interaction, mainly molecular weight, hydrogen bond formation, conformation, ionic attractions… Strong electrostatic interactions contribute to the force of bioadhesion.

At the surface of the platelets, chitosan chains are highly stretched and concentrated. The area of contact of a platelet with a biological surface is large, by comparison to a spherical device. Therefore, the adhesion force of a platelet is much larger and helps chitosan chains to penetrate into mucin reach domains. The electrostatic forces involving chitosan and mucin chains may stretch and pull the outer layer of chitosan of the platelets. If the cohesion force of chitosan inside the platelet is less than the adhesive force between chitosan and mucin, the system breakdown may occur in the platelet interlayer. The superficial layer of might be teared off the platelet and bind to the mucosa for instance, while the platelet comes apart. The "renewed" platelet can migrate to another site and the mechanism can start again. Layer after layer the platelet can possibly breakdown and provide a large coverage of the exposed mucosa through this mechanism. A combined effect can also occur: if the grafted chains may contribute to some therapeutic effect, they become available layer after layer. The competition between mucoadhesion interactions and platelet cohesion may control the behavior of the platelets used as a drug delivery device.

The literature shows that nanoparticles with functionalized $\beta$-CDs deliver non-soluble drugs for many therapeutic applications. If one could incorporate $\beta$-CDs in platelets in the same way as the $\alpha$-CDs, then the platelets can exhibit enhanced properties for drug delivery. Substitution of the inclusions occurred in our platelets and this property can provide an additional mechanism of drug release from platelets with therapeutic benefits.

## 4. CONCLUSION

This paper presents an extensive investigation of the structure and thermal properties of $\alpha$-CD inclusion complexes built from alkyl-grafted polysaccharides, with various lengths of alkyl chains. In all preparations, the inclusion complexes formed crystals with a platelet shape with nanometric thickness and high aspect ratio. All crystals contain at equilibrium water molecules in their structures with a proportion increasing with ambient relative humidity and showing specific thermal stability. All humid platelets exhibit identical X-Ray diffraction patterns, which belong to the "generic" channel type structure. The humid samples have a monoclinic symmetry and their unit cell volume is around 4500 $\text{Å}^3$. According to Harata and Kawano [19] the symmetry corresponds to the head to tail packing mode of $\alpha$-CD molecules, with the molecular axis inclined against the column axis. The platelets also contain the backbones of the polymer chains. SEM FEG images show that the platelets are composed of successive thin layers where the $\alpha$-CD rich inclusions and polysaccharide backbone could alternate. None of the crystalline structures reported so far in the literature of $\alpha$-CD inclusions shows such structures. Upon dehydration, platelets keep their crystalline structure with a decrease of the unit cell volume. Solubility and melting temperatures of platelets differ from pure crystals of $\alpha$-CD. The platelets in aqueous solutions are remarkably stable against temperature and show a cooperative melting temperature above 100 °C, for the palmitoyl-grafted polysaccharides, depending on the polymer backbone. One can envisage new drug delivery applications of platelets in relation with their specific structure. The crystalline platelets are susceptible to disintegrate by interacting with other molecules of their environment following a mechanism specific to these auto assemblies.

**Supporting Information Description:** The Supporting information summarizes the compositions of the platelets with different polysaccharides and grafted alkyl chains and provides additional data about the X-Ray diffraction patterns and their interpretation using the software Highscore Plus 3.0®.

## Acknowledgments

Authors acknowledge S. Suffit for technical support during SEM-FEG observations performed in the clean-room of the Laboratoire Matériaux et Phénomènes Quantiques (UMR 7162) at the Université Paris Diderot. The authors thank the reviewer of the manuscript for the careful examination and for useful comments and suggestions.

## TOC GRAPHIC

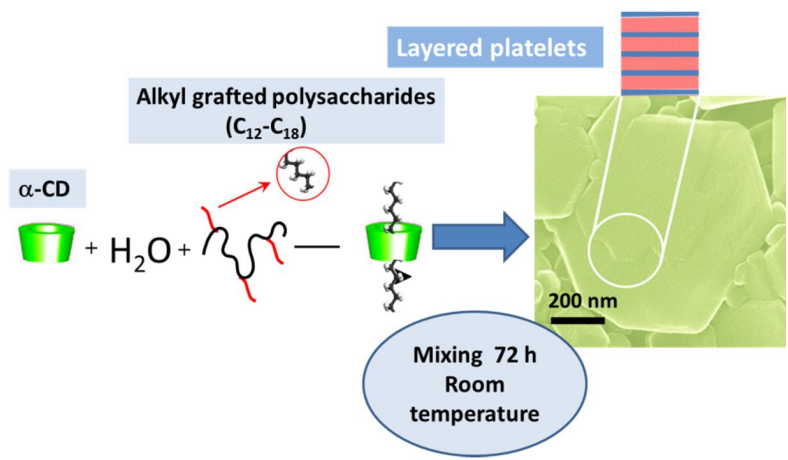

Layered platelets

Alkyl grafted polysaccharides
($C_{12}$-$C_{18}$)

α-CD

+ $H_2O$ +

Mixing  72 h
Room
temperature

200 nm